\documentclass[aps,twocolumn,pre]{revtex4}
\usepackage{graphicx}
\usepackage{amsfonts}
\usepackage{amssymb}
\usepackage{amsmath}

\begin{document}

\title{\bf Oscillations of dark solitons in trapped Bose-Einstein condensates}

\author{Dmitry E. Pelinovsky$^{1}$, D.J. Frantzeskakis$^{2}$, and P.G. Kevrekidis$^{3}$}

\affiliation{
$^{1}$ Department of Mathematics, McMaster University, Hamilton, Ontario, Canada, L8S 4K1 \\
$^{2}$ Department of Physics, University of Athens, Panepistimiopolis, Zografos, Athens 15784, Greece \\
$^{3}$ Department of Mathematics and Statistics, University of Massachusetts, Amherst MA 01003-4515, USA}

\begin{abstract}
We consider a one-dimensional defocusing Gross--Pitaevskii equation
with a parabolic potential. Dark solitons oscillate near the center of
the potential trap and their amplitude decays  due to radiative
losses (sound emission). We develop a systematic asymptotic
multi-scale expansion method in the limit when the potential trap is
flat. The first-order approximation predicts a uniform frequency of
oscillations for the dark soliton of arbitrary amplitude. The
second-order approximation predicts the nonlinear growth rate of the
oscillation amplitude, which results in decay of the dark soliton.
The results are compared with the previous publications and
numerical computations.
\end{abstract}

\maketitle

\section{Introduction}

Dark matter-wave solitons in atomic Bose--Einstein condensates
(BECs) have many similarities with dark optical solitons in
defocusing nonlinear media \cite{KL98,PPFA04}. Both entities are
fundamental nonlinear excitations of the defocusing nonlinear
Schr\"{o}dinger (NLS) equation, describing the evolution of electric
field envelopes in the context of optics, or of the order parameter
(the condensate mean-field wavefunction) in the context of atomic
BECs. Dark matter-wave solitons were observed in a series of
experiments carried out with BECs confined in external parabolic
magnetic trapping potentials \cite{exp1,exp2,exp3}, and have
inspired subsequent investigations of their stability and dynamics.
In particular, if a dark (black) soliton is initially placed exactly
at the center of the magnetic trap, it remains standing, while if it
is misplaced, it starts oscillating near the center of the trap. In
this respect, there has been a recent interest in theoretical
studies concerning the frequency and amplitude of these
oscillations, and especially on their dependence on the velocity and
amplitude parameters of dark solitons.

Preliminary studies of small-amplitude oscillations of dark solitons
were reported in \cite{RC97,MBB97} by using the collective
coordinate approach and assumptions of the soliton's adiabatic
dynamics. The frequency of oscillations obtained in
\cite{RC97,MBB97} was found to mismatch the correct frequency as
explained in \cite{BA00,FT02,BK03}. The main reason for discrepancy
of results of \cite{RC97,MBB97} is the use of the center-of-mass
quantity (the Ehrenfest Theorem), as the integral quantity can
change due to radiation from dark soliton \cite{BK03}. Another
integral quantity (the renormalized power invariant) was used in
\cite{BK03} (based on results of \cite{KV94}), but the perturbation
theory led to many correction terms in the main equations and
relatively poor agreement with numerical simulations (see, e.g.,
Fig. 1 in \cite{BK03}).

Another version of the perturbation theory for dark solitons can be
developed with the use of the renormalized momentum \cite{KY94} and
can be adapted to incorporate radiative effects on the adiabatic
dynamics of dark solitons \cite{PKA96}. This version of the
perturbation theory relies on the completeness of eigenfunctions of
the linearization problem established in \cite{CCH98,HCC99}. A recent
application of the perturbation theory, also accounting for radiative
effects, has been reported in \cite{L04} for dark solitons in the presence of linear
gain and two-photon absorption. The frequency of small-amplitude
oscillations of dark solitons near the center of the trapping
potential (and additional localized impurities) can be found from an integration
of the renormalized momentum \cite{FT02} in a much simpler form, as
compared to the one presented in \cite{BK03}.

Other integral invariants have also been used to study the adiabatic dynamics
of dark solitons, e.g. the boundary-layer integral for the
corresponding hydrodynamic equations \cite{BA00} and the energy
(Hamiltonian) of the dark soliton \cite{KP04}. Oscillations of dark
solitons were also studied in the shallow soliton limit with the use
of the Korteweg--de Vries (KdV) approximation \cite{HH02}. Surprisingly enough, it was found
that the frequency of oscillations of dark solitons does not depend on the soliton amplitude
\cite{KP04}, or, in other words, the frequency of oscillations remained the same in the limits of
black and shallow solitons \cite{BK03,HH02}.

Recent numerical studies of oscillations of dark solitons in
parabolic and other external traps were reported in
\cite{prok1,prok2,prok3}. It was found that the dark solitons emit
radiation (in the form of sound waves) due to oscillations. If
radiation escapes the trap (as in the case of a tight dimple trap
\cite{prok1}, or in an optical lattice potential \cite{prok3}), the
energy (momentum) of the dark soliton changes, resulting in the
growth of the oscillation amplitude and the decay of the soliton
amplitude. A phenomenological explanation of the radiative decay of
the soliton energy and the quadratic dependence of the energy decay
rate on the soliton acceleration was proposed in \cite{prok1,prok2},
based on the earlier analysis of \cite{PKA96}. However, the authors
of \cite{PKA96} consider instability-induced dynamics of dark
solitons in a homogeneous system, which is a different problem from
dynamics of dark solitons in a trapped condensate.

Further variants of the problem include oscillations of ring dark
solitons and vortex necklaces in two dimensions (i.e., for
``pancake'' BECs) \cite{TF03}, dynamics of shallow dark solitons in
a gas of hard core bosons (the so-called Tonks-Girardeau gas which,
in mean-field picture, is described by a quintic nonlinear
Schr\"{o}dinger equation) \cite{FPK04}, parametric driving of dark
solitons by periodically modulated Gaussian paddles \cite{PP04}, and
scattering of a dark soliton on a finite size obstacle \cite{BP05}.
It should be mentioned that the technique presented in \cite{PP04}
may pave the way for observing long-lived oscillating dark
matter-wave solitons in future BEC experiments.

The present paper deals with the oscillations of a dark soliton in a
BEC confined in a parabolic trap, as well as the
inhomogeneity-induced emission of radiation. Our analytical approach
relies on a systematic asymptotic multi-scale expansion method,
based on ideas of \cite{KY94,PKA96}, as well as the perturbation
theory for dark solitons \cite{CCH98,HCC99}. We prove that the main
equation of motion for adiabatic dynamics of dark solitons of any
amplitude is given by the harmonic oscillator equation with a
constant frequency. Additionally, we account for radiation escaping
the dark soliton and compute the nonlinear growth rate of the
oscillation amplitude versus the amplitude of a dark soliton. Our
results give a systematic basis for analysis of the dynamics of dark
solitons in other systems. The analytical results are found to be
very similar to the ones reported in recent studies
\cite{KP04,prok3}, except that a regular asymptotic technique, based
on the explicit small parameter of the problem, replaces previous
qualitative estimates based on numerical observations. Furthermore,
we show why a formal application of the perturbation theory fails to
incorporate the correct dependence of frequency of dark soliton
oscillations.

The paper is organized as follows. Section 2 formulates the problem
and reports the main results. Section 3 describes the asymptotic
limit for the ground state of the parabolic potential. Section 4
gives a transformation of the Gross-Pitaevskii (GP) equation to the
regularly perturbed nonlinear Schr\"{o}dinger (NLS) equation.
Section 5 contains the analysis of the perturbed NLS equation up to
the first-order and second-order corrections. Section 6 reviews the
formal application of perturbation theory to the GP equation and the
previous results. Section 7 concludes the paper.

\section{Model and main results}

At low temperatures, the dynamics of a repulsive quasi-one-dimensional BEC, oriented along the $x$--axis,
can be descibed by the following effective one-dimensional (1D) GP equation (see, e.g., \cite{gp1d})
\begin{equation}
i \hbar \psi_{t} = - \frac{\hbar^{2}}{2m} \psi_{xx} + V(x)\psi + g |\psi|^{2} \psi,
\label{dimgpe}
\end{equation}
where subscripts denote partial derivatives, $\psi(x,t)$ is the
mean-field BEC wavefunction, $m$ is the atomic mass, and the
nonlinearity coefficient $g$ (accounting for the interatomic
interactions) has an effective 1D form, namely $g=2 \hbar a
\omega_{\perp}$, where $a$ is the s-wave scattering length and
$\omega_{\perp}$ is the transverse-confinement frequency.
Additionally, the external potential $V(x)$ is assumed to be the
usual harmonic trap, i.e., $V(x) = m \omega_{x}^{2} x^2/2$, where
$\omega _{x}$ is the confining frequency in the axial direction.

To reduce the original GP equation (\ref{dimgpe}) to a dimensionless
form, $x$ is scaled in units of the fluid healing length
$\xi=\hbar/\sqrt{n_{0} g m}$ (which also characterizes the width of
the dark soliton), $t$ in units of $\xi/c$ (where
$c=\sqrt{n_{0}g/m}$ is the Bogoliubov speed of sound), the atomic
density $n \equiv |\psi|^{2}$ is rescaled by the peak density
$n_{0}$, and energy is measured in units of the chemical potential
of the system $\mu=g n_{0}$. This way, the following normalized GP
equation is readily obtained,
\begin{equation}
\label{GP} i u_t = - \frac{1}{2} u_{xx} + \epsilon^2 x^2 u + |u|^2
u,
\end{equation}
where the parameter $\epsilon \equiv (2\sqrt{2}a
n_{0})^{-1}(\omega_{x}/\omega_{\perp})$ determines the magnetic trap
strength and $u(x,t) \in \mathbb{C}$. Let us assume realistic
experimental parameters for a quasi-1D repulsive condensate
containing $N \sim 10^{3}$--$10^4$ atoms and with peak atomic
density $n_{0} \approx 10^8$ m$^{-1}$. Then, as the scattering
length $a$ is of order of a nanometer (e.g., $a=5.8$ nm or $a=2.7$
nm for a $^{87}$Rb or $^{23}$Na condensate), and the ratio of the
confining frequencies (for such a quasi-1D setting) is
$\omega_{x}/\omega_{\perp} \sim 1/200$, it turns out that the
magnetic trap strength $\epsilon$ is typically O$(10^{-2})$. Thus,
$\epsilon$ is a natural small parameter of the problem.

When $\epsilon = 0$, the defocusing GP
equation (\ref{GP}) has the exact solution for the dark soliton:
\begin{equation}
\label{ds} u_{\rm ds}(x,t) = \left[ k \tanh(k(x - vt - s_0)) + i v
\right] e^{-i t + i \theta_0},
\end{equation}
where $k = \sqrt{1 - v^2} < 1$ is the amplitude of the dark soliton
(with respect to the continuous-wave background), $|v| < 1$ is the
velocity parameter, and $(s_0,\theta_0) \in \mathbb{R}^2$ are
arbitrary parameters of the position and phase. Since
$$
|u_{\rm ds}|^2 = 1 - k^2 {\rm sech}^2(k(x-vt-s)),
$$
it is clear that the continuous-wave background for the dark soliton
(or the dimensionless chemical potential $\mu_{0}$) is normalized by
one, such that $\lim_{|x| \to \infty} |u_{\rm ds}|^2 = 1$. When $k
\to 1$ and $|v| \to 0$, the dark soliton approaches the limit of a
standing topological soliton (called the black soliton). When $k \to
0$ and $|v| \to 1$, the dark soliton approaches the limit of a
small-amplitude shallow soliton (which satisfies the KdV
approximation \cite{HH02}).

It should be noticed that, in physical terms, the choice $\mu_{0}=1$
actually sets the number of atoms $N$ of the condensate. In the
framework of the Thomas-Fermi approximation \cite{dalfovo}, it can
be found that $N=(4\sqrt{2}/3) (\xi n_{0}/\Omega) \mu_{0}^{3/2}$,
and, thus, for $n_{0} \approx 10^8$ m$^{-1}$, $\xi \sim 0.1$--$1$
microns (which are realistic value of the healing length for
quasi-1D Rb BECs) and $\Omega \sim 10^{-2}$, the choice $\mu_{0}=1$
leads to a number of atoms of the order of $N \sim 10^3$--$10^4$.

When $\epsilon \neq 0$, but the nonlinearity is crossed out by the
linearization, the parabolic potential of the GP equation (\ref{GP})
has the ground state solution:
\begin{equation}
\label{gs} u_{\rm gs}(x,t) = u_0 \exp\left(-\frac{\epsilon x^2 + i
\epsilon t}{\sqrt{2}}\right),
\end{equation}
where $u_0 \in \mathbb{C}$ is an arbitrary parameter of the ground
state amplitude. When the nonlinear term of the defocusing GP
equation (\ref{GP}) is taken into account, the linear mode
(\ref{gs}) generates a family of ground state solutions by means of
a standard local bifurcation \cite{kunze},
\begin{equation}
\label{gs-main} u_{\rm gs}(x,t) = U_{\epsilon}(x) e^{-i
\mu_{\epsilon} t + i \theta_0},
\end{equation}
where $\mu_{\epsilon} \in \mathbb{R}$ is the normalized chemical
potential and $\theta_0 \in \mathbb{R}$ is an arbitrary phase. Due
to the scaling invariance of the GP equation (\ref{GP}), the
amplitude of the ground state $U_{\epsilon}(x) \in \mathbb{R}$ can
be uniquely normalized by one, such that $|U_{\epsilon}(0)|^2 = 1$.

The first excited state of the parabolic potential bifurcates from
the linear solution:
\begin{equation}
\label{first-excited-state} u_{1es}(x,t) = x
\exp\left(-\frac{\epsilon x^2 + 3 i \epsilon t}{\sqrt{2}} \right),
\end{equation}
by means of the same local bifurcation \cite{kunze}. The first
excited state corresponds to a static bound state between the dark
soliton (\ref{ds}) placed at the center $x = 0$ of the nonlinear
ground state (\ref{gs-main}). The solution for the static bound
state exists for any $\epsilon \neq 0$ but tells nothing about
dynamics of the dark soliton placed near the center of the ground
state.

Dynamics of dark soliton is considered in this paper. We show that
the dark soliton (\ref{ds}) undertakes adiabatic dynamics in the
limit $\epsilon \to 0$, such that the parameter $(s_0 + vt) \equiv
s(T)/\epsilon$ of position of the dark soliton (\ref{ds}) becomes a
function of slow time $T = \epsilon t$, while the velocity parameter
$v$ is defined as $v(T) = \dot{s}$. The adiabatic dynamics leads to
generation of radiative waves, which escape the dark soliton but
become trapped by the parabolic potential. In the decomposition of
the solution $u(x,t)$ into two (inner and outer) asymptotic scales,
the leading-order radiative effects are taken into account when
parameter $\theta_0 \equiv \theta(T)$ of complex phase of the dark
soliton (\ref{ds}) depends also on $T = \epsilon t$ and the
first-order corrections to the dark soliton (\ref{ds}) grow linearly
in $x$. When reflections from the trapping potential are neglected,
the extended dynamical equation for the position $s(T)$ of the dark
soliton (\ref{ds}) takes the form:
\begin{equation}
\label{pumped-oscillator-main} \ddot{s} + s = \frac{\epsilon
\dot{s}}{2 \sqrt{(1-s^2)^3} \sqrt{1 - s^2 - \dot{s}^2}} + {\rm
O}(\epsilon^2),
\end{equation}
in the domain $(s,\dot{s}) \in {\cal D}_0$, where ${\cal D}_0$ is
the unit disk:
\begin{equation}
\label{disc} {\cal D}_0 = \{ (s,\dot{s}) \in \mathbb{R}^2 : \;\;
s^2 + \dot{s}^2 < 1 \}.
\end{equation}
The left-hand-side of the dynamical equation
(\ref{pumped-oscillator-main}) represents the leading-order
adiabatic dynamics of the dark soliton oscillating on the ground
state of the trapping potential. The right-hand-side represents the
leading-order radiative effects (sound emission), when reflections
of radiation from the parabolic potential are neglected. Within the
truncation error of ${\rm O}(\epsilon^2)$, the ground state
(\ref{gs-main}) can be approximated by the Thomas--Fermi (TF)
approximation $U_{\epsilon}(x) = \sqrt{1 - \epsilon^2 x^2}$ on the
scale $|x| < \epsilon^{-1}$.

The only equilibrium point of the dynamical equation
(\ref{pumped-oscillator-main}) is $(0,0)$ and it corresponds to the
static bound state bifurcating from the first excited state
(\ref{first-excited-state}). The leading-order part of the dynamical
equation (\ref{pumped-oscillator-main}) describes a harmonic
oscillator with the obvious solution: $s(T) = s_0 \cos(T +
\delta_0)$. This result is well-known from earlier papers
\cite{BA00,FT02,BK03}, where it was derived in the limit of black
soliton, when $k \to 1$ and $|v| \to 0$. In the derivation presented
herein, we have not used the assumption on the initial position
$s(0)$ and speed $\dot{s}(0)$ of the dark soliton, and therefore,
the approximation of the harmonic oscillator
(\ref{pumped-oscillator-main}) remains valid for larger values of
$(s,\dot{s})$ inside the unit disk (\ref{disc}).

We note that the velocity-dependent correction to the frequency of
the harmonic oscillator (\ref{pumped-oscillator-main}) was obtained in an earlier study
(see Eq. (36) in \cite{BK03}), but it is not confirmed within our
analysis. On the other hand, our results confirm the results of the
shallow soliton approximation \cite{HH02} which establishes the same
frequency of oscillations as in the black soliton limit. In
addition, the uniform frequency of oscillations for dark solitons of
all amplitudes and velocities was recently reported by means of the
energy (Hamiltonian) computations \cite{KP04}.

Let $E$ be the energy of the harmonic oscillator:
\begin{equation}
E = \frac{1}{2} \left( \dot{s}^2 + s^2 \right).
\end{equation}
The energy increases in time due to the first-order correction terms
of the main equation (\ref{pumped-oscillator-main}):
\begin{equation}
\label{energy-increase} \dot{E} = \frac{\epsilon \dot{s}^2}{2
\sqrt{(1-s^2)^3} \sqrt{1 - s^2 - \dot{s}^2}} + {\rm O}(\epsilon^2)
> 0,
\end{equation}
where $(s, \dot{s}) \in {\cal D}_0$. Due to the energy pumping
(\ref{energy-increase}), the amplitude of the harmonic oscillator
increases in time. When the oscillation amplitude grows, the dark
soliton (\ref{ds}) shifts from the black soliton limit $k \to 1$,
$|v| \to 0$ to the shallow soliton limit $k \to 0$, $|v| \to 1$. By
the Poincar\'{e}-Bendixon Theorem, the limit cycle does not exist in
the unit disk (\ref{disc}) and all orbits approach the boundary of
the disk, where the main equation (\ref{pumped-oscillator-main})
becomes invalid. The growth rate of the oscillation amplitude is
nonlinear in general and depends on $(s,\dot{s})$.

In the limit $s^2 + \dot{s}^2 \to 0$, the main equations
(\ref{pumped-oscillator-main}) and (\ref{energy-increase}) can be
simplified. First, the energy of the dark soliton oscillations
accelerates by the squared law $\dot{E} = \epsilon \dot{s}^{2}/2$,
which is postulated in \cite{prok1} and confirmed in numerical
computations where the dark solitons oscillated in a tight dimple
trap (see Fig. 2 in \cite{prok1}). Second, the nonlinear equation
(\ref{pumped-oscillator-main}) is linearized as follows:
$$
\ddot{s} + s - \frac{\epsilon}{2} \dot{s} = {\rm O}(\epsilon^2,
s^3),
$$
which shows that the center point $(0,0)$ becomes an unstable spiral
point on the plane $(s,\dot{s})$ as $0 < \epsilon \ll 1$, with the
leading-order solution: $s(T) = s_0 e^{\epsilon T/4} \cos(T +
\delta_0)$. Therefore, due to radiative losses, the amplitude of
oscillations of dark solitons increases while its own amplitude
decreases. This main result of the asymptotic analysis was also
confirmed by numerical simulations where the dark solitons
oscillated between two Gaussian humps (see Fig. 9 in \cite{prok2}).

\begin{figure}[tbp]
\includegraphics[width=8cm]{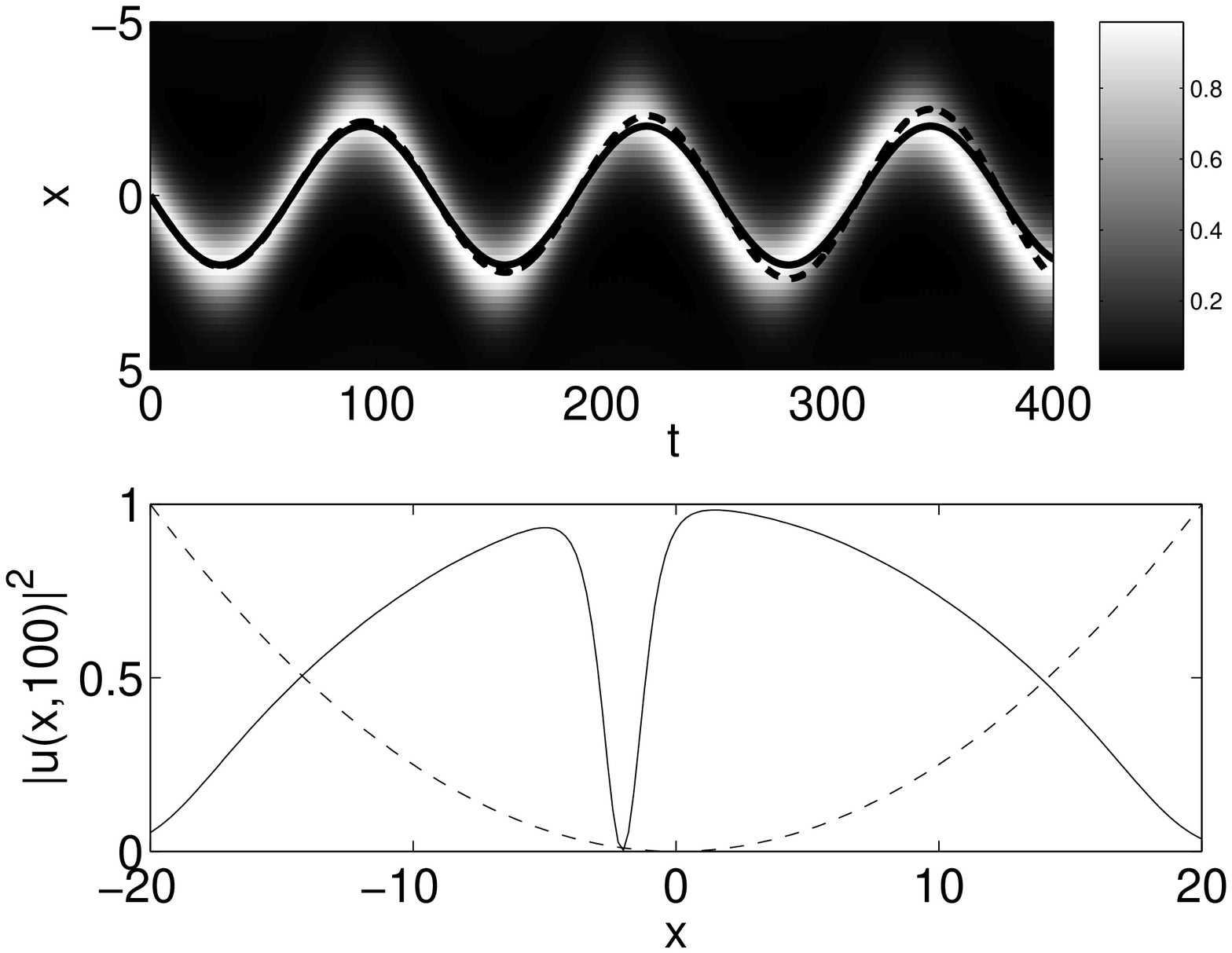}
\includegraphics[width=8cm]{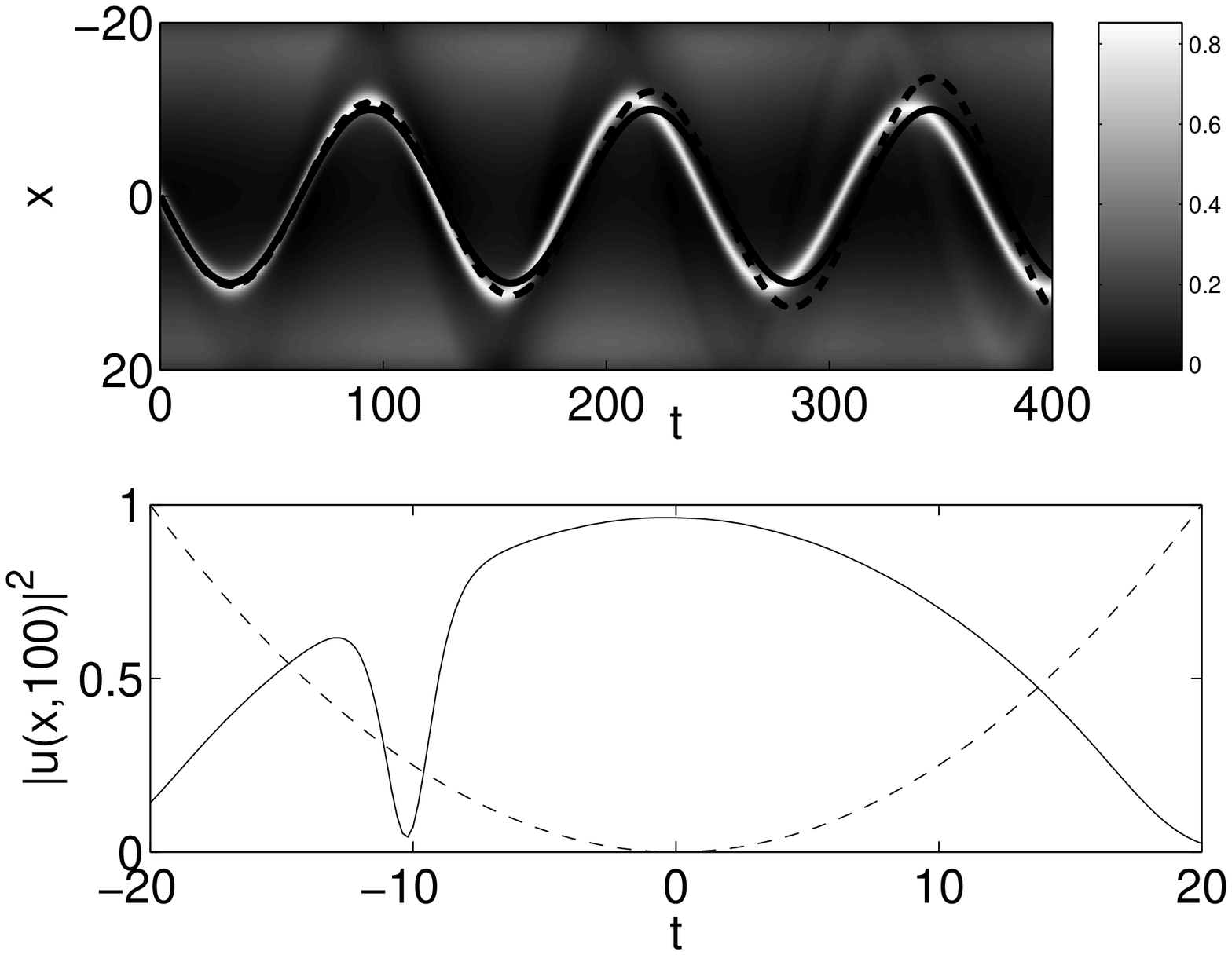}
\caption{Spatio-temporal evolution of the reduced condensate density
(the ground state density minus the actual density) for the GP
equation (\ref{GP}) with $\epsilon = 0.05$, $s(0) = 0$, and $v(0) =
0.1$ (top two panels), $v(0) = 0.5$ (bottom two panels). The solid lines
correspond to the solutions of the harmonic (adiabatic)
approximation and the dashed lines correspond to the anti-damped
approximation (pertaining to sound emission). The dotted line on the
plot of $|u(x,100)|^2$ shows the parabolic trapping potential.}
\label{fig1}
\end{figure}

Figure \ref{fig1} shows a spatio-temporal contour plot of the
reduced density (the ground state density subtracted from the actual
density) for a one-dimensional BEC confined in a parabolic trap with
normalized strength $\epsilon=0.05$ (the TF radius is equal to
$20$). The figures were obtained by numerical integration of the GP
equation (\ref{GP}). The white areas correspond to a dark soliton,
initially placed at the trap center (i.e., $s(0)=0$) with an initial
velocity $v(0) = 0.1$ (bottom panel) and $v(0)=0.5$ (top panel). The
plot is compared to the two versions of the main equation
(\ref{pumped-oscillator-main}). The solid lines correspond to
harmonic oscillations (within the adiabatic soliton dynamics), while
the dashed ones correspond to the growing oscillations (within the
inhomogeneity-induced sound emission). It is seen from the figures
that the dashed lines approximate better the actual dark soliton
motion within the first period of oscillations for $0 \leq t \leq
t_* < T_{\rm osc}$, where $T_{osc} = 2\pi/\epsilon$. For instance, a
better agreement is achieved at the first turning points where, due
to stronger sound emission, a slight increase of the amplitude of
soliton oscillations is readily observed.

Nevertheless, for longer times, the anti-damped approximation of the
main equation (\ref{pumped-oscillator-main}) becomes irrelevant due
to the fact that the radiation cannot escape the trap and is
reflected back to the dark soliton. As a result, the soliton
continuously interacts with the reflected radiation so that, on
average, the soliton reabsorbs the radiation it emits on the
contrast to numerical computations in \cite{prok1,prok2}. We note
that the sound emission is much weaker for the deeper soliton (with
$v=0.1$) and consequently the analytical result pertaining to the
adiabatic approximation of the soliton motion is much closer to the
result obtained by the numerical simulation.

The developed method allows us to incorporate multiple sound
reflections and their recombination to the asymptotic equations for
the dark soliton. If this is done, the main equation
(\ref{pumped-oscillator-main}) is extended by an additional term due
to multiple reflections beyond the initial time interval $0 \leq t
\leq t_*$. This complication is however beyond the scopes of the
present paper.

\section{Ground state of the GP equation}

We start with analysis of the ground state solution of the GP
equation (\ref{GP}) in the form (\ref{gs-main}), where
$(U_{\epsilon}(x),\mu_{\epsilon})$ is a real-valued pair of
eigenfunctions and eigenvalues of the nonlinear boundary-value
problem:
\begin{equation}
\label{boundary-value} \frac{1}{2} U'' - \epsilon^2 x^2 U - U^3 +
\mu U = 0, \qquad x \in \mathbb{R}^+,
\end{equation}
subject to the normalized boundary conditions:
\begin{equation}
\label{ground-state}  U_{\epsilon}(0) = 1, \quad U_{\epsilon}'(0) =
0, \quad \lim_{x \to \infty} U_{\epsilon}(x) = 0.
\end{equation}
We are interested in existence of the symmetric ground state
$U_{\epsilon}(x) > 0$, $U_{\epsilon}(-x) = U_{\epsilon}(x)$ on $x
\in \mathbb{R}$ for small values of $\epsilon^2$. We recall two
facts for the boundary-value problem (\ref{boundary-value}) with a
given value of $\epsilon > 0$ (see \cite{kunze}): (i) the local
bifurcation from the linear ground state (\ref{gs}) occurs for $\mu
> \mu_0(\epsilon)$, where $\mu_0 = \epsilon/\sqrt{2}$,
and (ii) the nonlinear ground state (\ref{gs-main}) exists as a
one-parameter smooth family, parameterized by $\mu$ or,
equivalently, by $U(0)$. Moreover, $U(0)$ is an increasing function
of $\mu > \mu_0(\epsilon)$ for a fixed value of $\epsilon$.
Therefore, for a given value of $\epsilon > 0$, there exists a
unique value of $\mu$ (called $\mu_{\epsilon}$), which corresponds
to the normalization $U_{\epsilon}(0) = 1$ and the solution pair
$(U_{\epsilon}(x),\mu_{\epsilon})$ is a smooth function of
$\epsilon$. The solution of the boundary-value problem of
(\ref{boundary-value})--(\ref{ground-state}) can be approximated
numerically by means of a contraction mapping method. Figure 2 shows
the dependence of $\mu_{\epsilon}$ versus $\epsilon$, the profile
$U_{\epsilon}(x)$ for different values of $\epsilon$, and the
dependence of $U(0)$ versus $\mu$, for fixed $\epsilon = 0.05$.

\begin{figure}[tbp]
\includegraphics[width=8cm]{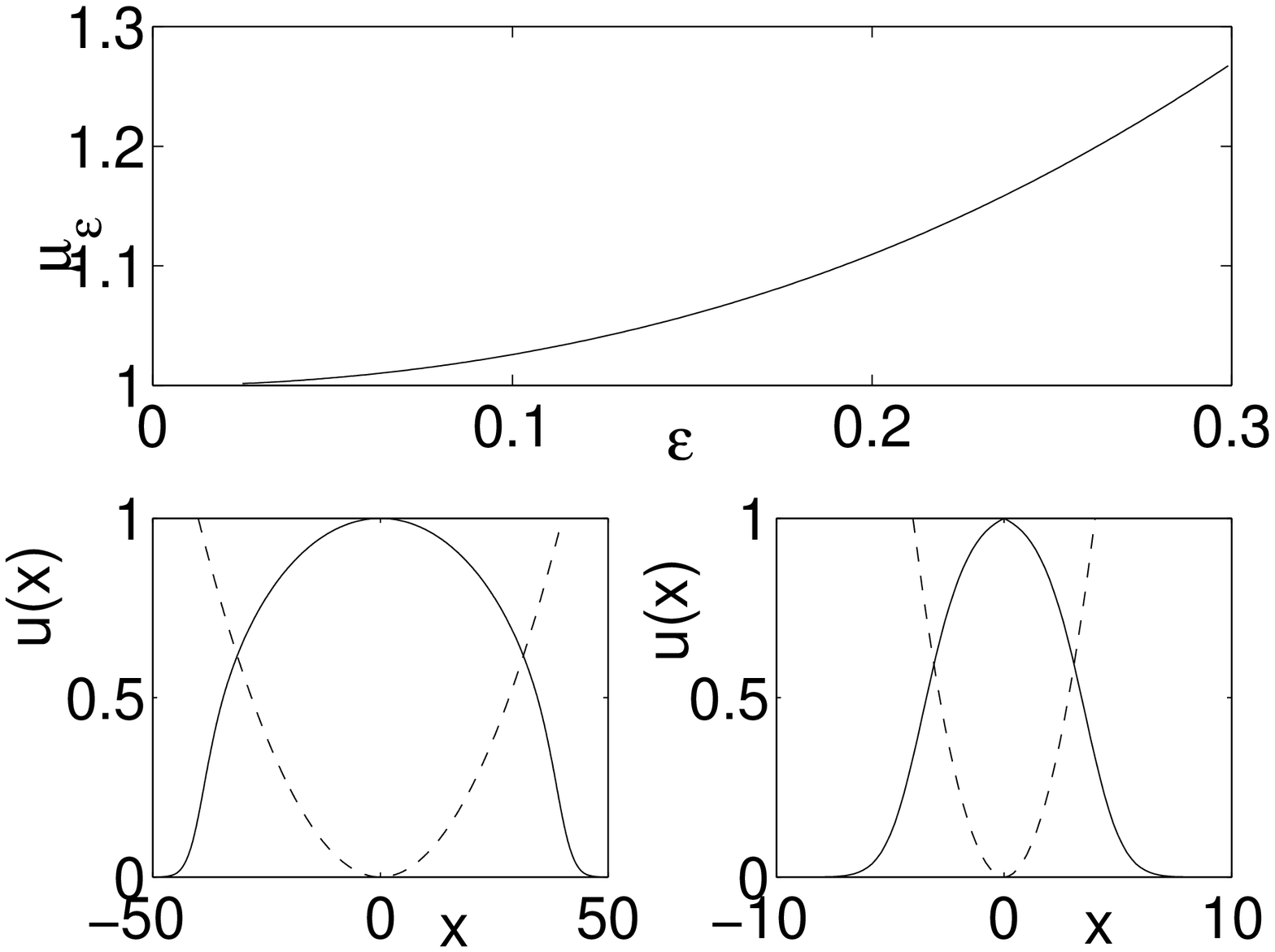}\\
\includegraphics[width=7cm]{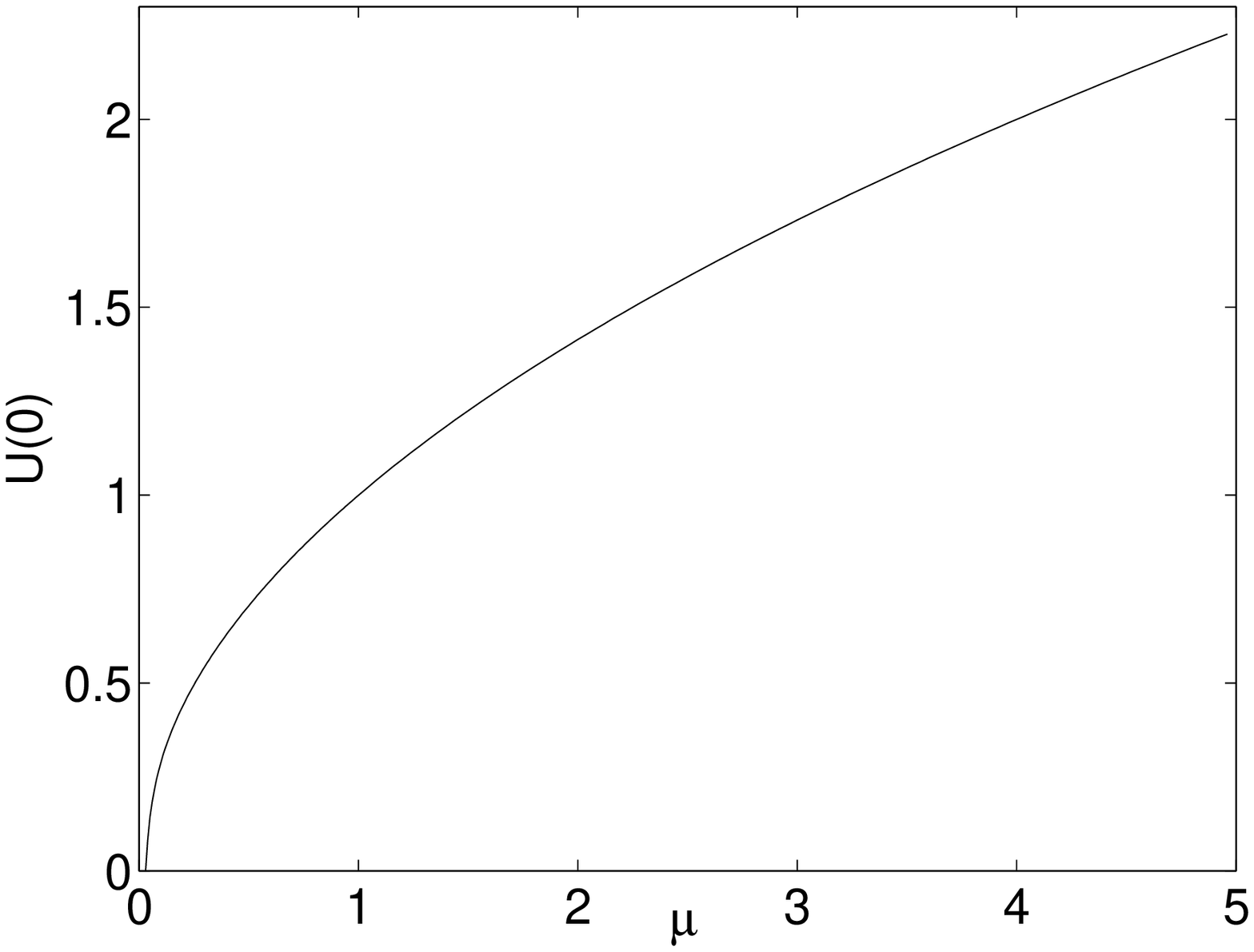}
\caption{Ground state solution of the boundary-value problem
(\ref{boundary-value}). Top panel: The dependence of
$\mu_{\epsilon}$ versus $\epsilon$ for fixed $U_{\epsilon}(0)=1$.
Middle panels: Profiles of the function $U(x)$ for two different
values of the trap strength: $\epsilon=0.025$ (left) and
$\epsilon=0.25$ (right). Bottom panel: The dependence of $U(0)$
versus $\mu$ for fixed $\epsilon=0.05$.} \label{fig2}
\end{figure}

The solution of the ODE (\ref{boundary-value}) with initial values
$U(0) = 1$ and $U'(0) = 0$ can be constructed in the power series form:
\begin{equation}
\label{power-series} U(x) = 1 + \sum_{k=1}^{\infty} a_k x^{2k},
\end{equation}
where coefficients $\{ a_k \}_{k=1}^{\infty}$ can be found
recursively, e.g. the first two terms are:
\begin{eqnarray*}
a_1 & = & 1 - \mu, \\
a_2 & = & \frac{1}{6} \left( \epsilon^2 + (1-\mu) (3-\mu) \right).
\end{eqnarray*}
The parameter $\mu = \mu_{\epsilon}$ is defined from the decay
condition at infinity, where $U = U_{\epsilon}(x)$ and $\lim_{x \to
\infty} U_{\epsilon}(x) = 0$. However, the decay condition is
computationally inefficient for approximations of the dependence of
$\mu_{\epsilon}$ versus $\epsilon$. A different (WKB) method was
used for approximations of the solution $U_{\epsilon}(x)$ and the
dependence $\mu_{\epsilon}$ in \cite{BK03,HH02}. The advantage of
the WKB method over the power series (\ref{power-series}) is that
the dependence $\mu_{\epsilon}$ is determined from the condition at
$x = 0$ rather than from the decay condition as $x \to \infty$.

In the WKB method, the solution of the ODE (\ref{boundary-value}) is
represented in the form $U(x) = \sqrt{Q(x)}$, where $Q(x)$ satisfies
the equivalent problem:
\begin{equation}
\label{WKB-problem} Q(x) = \mu - \epsilon^2 x^2 + \frac{Q''(x)}{4
Q(x)} - \frac{Q^{\prime 2}(x)}{8 Q^2(x)}.
\end{equation}
In the limit of small $\epsilon$, the solution of the ODE
(\ref{WKB-problem}) can be thought in the form of a WKB asymptotic
series:
\begin{equation}
\label{WKB-series} Q = \mu - X^2 + \sum_{k=1}^{\infty} \epsilon^{2k}
Q_k(X),
\end{equation}
where $X = \epsilon x$, $|X| < \sqrt{\mu}$, and the set $\{ Q_k(X)
\}_{k=1}^{\infty}$ can be found recursively, e.g. the first two
terms are:
$$
Q_1 = - \frac{\mu}{2(\mu - X^2)^2}
$$
and
$$
Q_2 = \frac{Q_1''(X)}{4(\mu-X^2)} + \frac{X Q_1'(X) + Q_1(X)}{2(\mu
- X^2)^2} + \frac{X^2 Q_1(X)}{(\mu - X^2)^3}.
$$
The symmetry in $X$ implies that the condition $Q'(0) = 0$ is
satisfied. The normalization condition $Q(0) = 1$ defines the power
approximation of the dependence $\mu = \mu_{\epsilon}$ versus
$\epsilon$:
\begin{equation}
Q(0) = \mu - \frac{\epsilon^2}{2 \mu} - \frac{3 \epsilon^4}{4 \mu^3}
+ {\rm O}(\epsilon^6) = 1,
\end{equation}
such that
\begin{equation}
\label{first-terms-mu} \mu_{\epsilon} = 1 + \frac{\epsilon^2}{2} +
\frac{\epsilon^4}{2} + {\rm O}(\epsilon^6).
\end{equation}
Using the leading-order approximation for $\mu_{\epsilon}$, we find
an asymptotic approximation for $U_{\epsilon}(x)$ from the WKB
asymptotic series (\ref{WKB-series}):
\begin{equation}
\label{first-terms-U} U_{\epsilon}(x) = \sqrt{1 - \epsilon^2 x^2} +
\epsilon^2 \tilde{U}(\epsilon x, \epsilon),
\end{equation}
where $\tilde{U}(\epsilon x,\epsilon)$ is the remainder term. The
leading-order approximation in (\ref{first-terms-U}) is referred to
as the Thomas--Fermi approximation \cite{dalfovo}. The WKB asymptotic series
(\ref{WKB-series}) gives a bounded approximation of the solution
pair $(U_{\epsilon}(x),\mu_{\epsilon})$ for $|\epsilon x| < \sqrt{\mu_{\epsilon}}$
but it becomes invalid at and beyond the
turning points at $|\epsilon x| \geq \sqrt{\mu_{\epsilon}}$
\cite{WKBtext}. Therefore, the Thomas-Fermi approximation
(\ref{first-terms-U}) is valid only for $|\epsilon x | < 1$ in the
limit of small $\epsilon$.

\section{Transformation of the GP equation}

We proceed with analysis of dynamics of the dark soliton on the
ground state of the parabolic potential. This problem can be studied
after the equivalent transformation of the GP equation (\ref{GP}):
\begin{equation}
u(x,t) = U_{\epsilon}(x) w(x,t) e^{- i \mu_{\epsilon} t},
\end{equation}
where $(U_{\epsilon}(x),\mu_{\epsilon})$ is the solution pair of the
boundary-value problem (\ref{boundary-value})--(\ref{ground-state})
and $w(x,t)$ is a new variable. The function $w(x,t)$ satisfies the
PDE problem:
\begin{equation}
\label{GP-modified} i w_t + \frac{1}{2} w_{xx} + U_{\epsilon}^2(x)
(1 - |w|^2) w = - \frac{U_{\epsilon}'(x)}{U_{\epsilon}(x)} w_x.
\end{equation}
We will use the fact that $U_{\epsilon}(x) \equiv U_{\epsilon}(X)$,
$X = \epsilon x$ in the asymptotic region $|\epsilon x| < 1$ as
$\epsilon \to 0$. Assuming that the dynamics of the dark soliton
occurs inside the asymptotic region $|\epsilon x | < 1$, we
introduce the traveling wave coordinate for the dark soliton:
\begin{equation}
\eta = x - \frac{s(T)}{\epsilon}, \qquad T = \epsilon t,
\end{equation}
where the dependence $s(T)$ is to be determined, and expand the
function $U_{\epsilon}(X)$ in the Taylor series near $X = s(T)$:
\begin{eqnarray}
&&U_{\epsilon}(X) = U_{\epsilon}(s(T) + \epsilon \eta)   \nonumber \\
&&= U_{\epsilon}(s) + \epsilon \eta U_{\epsilon}'(s) + \frac{1}{2}
\epsilon^2 \eta^2 U_{\epsilon}''(s) + {\rm O}((\epsilon \eta)^3).
\end{eqnarray}
As a result, the PDE problem (\ref{GP-modified}) takes the form of
the perturbed equation:
\begin{eqnarray*}
i w_t &-& i v w_{\eta} + \frac{1}{2} w_{\eta \eta} + U_{\epsilon}^2(s) (1 - |w|^2) w  \\
&=& - \epsilon \left(\frac{U_{\epsilon}'(s)}{U_{\epsilon}(s)} w_{\eta} + 2
U_{\epsilon}(s) U_{\epsilon}'(s) \eta (1 - |w|^2) w \right) \\
&-& \epsilon^2 \frac{(U_{\epsilon}''(s) U_{\epsilon}(s) -
U_{\epsilon}^{\prime 2}(s))}{U_{\epsilon}^2(s)} \eta w_{\eta} \\
&-& \epsilon^2 \eta^2 \left( U_{\epsilon}(s) U_{\epsilon}''(s) +
U_{\epsilon}^{\prime 2}(s)\right) (1 - |w|^2) w + {\rm
O}(\epsilon^3),
\end{eqnarray*}
where $v(T) = \dot{s}$ is the speed of the dark soliton. The
perturbed equation is simplified with the scaling transformation:
\begin{equation}
\label{scaling-transformation} z = \eta U_{\epsilon}(s(T)), \qquad
v(T) = \nu(T) U_{\epsilon}(s(T)),
\end{equation}
such that the wave function $w(z,t)$ satisfies the perturbed
defocusing NLS equation:
\begin{eqnarray}
\nonumber i w_t + U_{\epsilon}^2(s) \left[ - i \nu w_z + \frac{1}{2}
w_{zz} + (1 - |w|^2) w \right] \\
+ R(w,\bar{w}) = 0, \label{NLS}
\end{eqnarray}
where $R = \epsilon R_1 + \epsilon^2 R_2 + {\rm O}(\epsilon^3)$
are perturbation terms, with the first two of them being:
\begin{eqnarray*}
R_1 &=&  U_{\epsilon}'(s) \left( i \nu z w_z + w_z + 2 z (1 - |w|^2) w \right) \\
R_2 &=&  \frac{(U_{\epsilon}''(s) U_{\epsilon}(s) -
U_{\epsilon}^{\prime 2}(s))}{U_{\epsilon}^2(s)} z w_z \\
&+& \frac{(U_{\epsilon}(s) U_{\epsilon}''(s) + U_{\epsilon}^{\prime 2}(s))}{U_{\epsilon}^2(s)} z^2 (1 - |w|^2) w.
\end{eqnarray*}
Since $U_{\epsilon}(s) > 0$ for the ground state of the parabolic
potential, the perturbation terms of the equation (\ref{NLS}) are
regular for any $s \in \mathbb{R}$. However, since the
representation $U_{\epsilon}(x) \equiv U_{\epsilon}(X)$ is valid
only for $|X| < 1$ in the limit of small $\epsilon$, we consider the
perturbed NLS equation (\ref{NLS}) in the region where $|s| < 1$.
The leading-order part of the defocusing NLS equation (\ref{NLS})
(when $R(w,\bar{w}) = 0$) has the exact solution for the dark
soliton:
\begin{equation}
\label{dark-soliton} w(z,t) = w_0(z) = \kappa \tanh(\kappa z) + i
\nu,
\end{equation}
where $\kappa = \sqrt{1 - \nu^2}$ and $\nu(T) = \nu_0$ is a constant
speed. The time evolution of the dark soliton (\ref{dark-soliton})
in the case $R(w,\bar{w}) \neq 0$ is studied with a regular
perturbation theory for dark solitons \cite{KY94,PKA96,CCH98,HCC99}.

\section{Asymptotic multi-scale expansion method}

The solution to the perturbed NLS equation (\ref{NLS}) can be
obtained in the form of an asymptotic multi-scale expansion series
for the perturbed dark soliton (\ref{dark-soliton}) \cite{PKA96}:
\begin{eqnarray}
\label{asymptotic-series} w(z,t) = \left[ w_0 + \epsilon w_1 +
\epsilon^2 w_2 + {\rm O}(\epsilon^3) \right] e^{i \theta},
\end{eqnarray}
where $T = (\epsilon t,\epsilon^2 t,...)$, the function $w_0 =
w_0(z;T)$ is the dark soliton (\ref{dark-soliton}), while the
functions $w_1 = w_1(z,t;T)$ and $w_2 = w_2(z,t;T)$ solve the
inhomogeneous linear problems:
\begin{eqnarray}
\nonumber i \partial_t \sigma_3 {\bf w}_1 + U_{\epsilon}^2(s) {\cal
H} {\bf w}_1 = \dot{\theta} {\bf w}_0 - i
\partial_T \sigma_3 {\bf w}_0 \\
- {\bf R}_1(w_0,\bar{w}_0)
\label{first-order-problem}
\end{eqnarray}
and
\begin{eqnarray}
\label{second-order-problem} i \partial_t \sigma_3 {\bf w}_2 +
U_{\epsilon}^2(s) {\cal H} {\bf w}_2
 = \dot{\theta} {\bf w}_1 - i \partial_T \sigma_3 {\bf w}_1 \nonumber \\
 - {\bf N}_2(w_1,\bar{w}_1) - D {\bf R}_1 (w_0,\bar{w}_0) {\bf w}_1 - {\bf R}_2(w_0,\bar{w}_0).
\end{eqnarray}
Parameters $s(T)$ and $\theta(T)$ are to be determined from
solutions of the inhomogeneous problems, while $\nu(T) = \dot{s}/
U_{\epsilon}(s)$. In the problems (\ref{first-order-problem}) and
(\ref{second-order-problem}), we have introduced the notations:
${\bf w}_k$ and ${\bf R}_k$ for vectors $(w_k,\bar{w}_k)^T$, $k =
0,1,2$ and $(R_k,\bar{R}_k)^T$, $k = 1,2$, $D {\bf R}_1$ for the
Jacobian of ${\bf R}_1$, ${\bf N}_2$ for the quadratic terms from
the left-hand-side of the unperturbed NLS equation. The self-adjoint
linearization operator is
$$
{\cal H} = - i \nu \sigma_3 \partial_z + \sigma_0 \left( \frac{1}{2}
\partial^2_z + 1 \right) -
\left( \begin{array}{ccc} 2 |w_0|^2 & w_0^2 \\ \bar{w}_0^2 & 2
|w_0|^2 \end{array} \right).
$$
where $\sigma_0 = {\rm diag}(1,1)$ and $\sigma_3 = {\rm
diag}(1,-1)$. Analysis of the first-order problem
(\ref{first-order-problem}) predicts a leading-order equation for
$s(T)$, which characterizes oscillations of dark solitons near the
center of the parabolic potential. The first-order problem also
describes the leading-order radiation from the dark soliton to
infinity, related to the equation for $\theta(T)$. Analysis of the
second-order problem (\ref{second-order-problem}) predicts a
first-order correction to the equation for $s(T)$, which is induced
by the leading-order radiation. Due to radiation, the oscillation
amplitude grows in time so that the amplitude of the dark soliton
decreases. Derivation of all these results is divided into several
technical subsections.

\subsection{Linearization operator}

The self-adjoint operator ${\cal H}$ is defined on complete Hilbert
space $H^1(\mathbb{R},\mathbb{C}^2) \subset
L^2(\mathbb{R},\mathbb{C}^2)$. It has a non-empty kernel:
\begin{equation}
\label{kernel-H} {\cal H} {\bf w}_0'(z)  = {\bf 0},
\end{equation}
which is related to translational invariance of the unperturbed NLS
equation in $z$. Furthermore, the operator ${\cal H}$ has two
branches of the continuous spectrum:
\begin{equation}
\label{continuous-H} \sigma_{\rm ess}({\cal H}) = (-\infty,-2] \cup
(-\infty,0],
\end{equation}
such that the second branch intersects the kernel. The only bounded
non-decaying eigenvector for the zero eigenvalue, which belongs to
the continuous spectrum (\ref{continuous-H}), is related to the
gauge invariance of the NLS equation:
\begin{equation}
\label{bounded-H} {\cal H} \left( i \sigma_3 {\bf w}_0 \right) =
{\bf 0}.
\end{equation}
The homogeneous equation ${\cal H} {\bf w} = {\bf 0}$ supports the
decaying solution (\ref{kernel-H}), the bounded solution
(\ref{bounded-H}), a linearly growing solution, and an exponentially
growing solution. The linearly growing solution can be found
explicitly \cite{PKA96}:
\begin{equation}
\label{growing-H} {\cal H} \left( i \sigma_3 z {\bf w}_0 -
\partial_{\nu} {\bf w}_0 + \frac{3 \nu}{2 \kappa} \partial_{\kappa} {\bf
w}_0 \right) = {\bf 0}.
\end{equation}
Here and henceforth, it is convenient to consider $w_0(z)$ as a
function of two independent parameters $\kappa$ and $\nu$. The
relation $\kappa = \sqrt{1 - \nu^2}$ is used after evaluating the
partial derivatives of $w_0(z)$ in $\kappa$ and $\nu$.

The linearization operator $\sigma_3 {\cal H}$ has the kernel
(\ref{kernel-H}) and the generalized kernel:
\begin{equation}
\label{genkernel-H} \sigma_3 {\cal H} \left( \partial_{\nu} {\bf
w}_0 - \frac{\nu}{\kappa} \partial_{\kappa} {\bf w}_0 \right) = i
{\bf w}_0'(z).
\end{equation}
The spectrum of $\sigma_3 {\cal H}$ includes two branches of the
continuous spectrum:
\begin{equation}
\label{continuous-sigmaH} \sigma_{\rm ess}\left(\sigma_3 {\cal
H}\right) = \mathbb{R} \cup \mathbb{R}.
\end{equation}
The spectrum of $\sigma_3 {\cal H}$ consisting of the kernel
(\ref{kernel-H}), the generalized kernel (\ref{genkernel-H}) and the
two branches of the continuous spectrum (\ref{continuous-sigmaH}) is
complete in $H^1(\mathbb{R},\mathbb{C}^2)$ \cite{CCH98,HCC99}.
Nevertheless, since the kernel of ${\cal H}$ has also a bounded
non-decaying eigenvector (\ref{bounded-H}), we construct a bounded
non-decaying eigenvector from the nonhomogeneous problem:
\begin{equation}
\label{genkernel-bounded-H} \sigma_3 {\cal H} \left( \frac{1}{2
\kappa}  \partial_{\kappa} {\bf w}_0 \right) = i \left( i \sigma_3
{\bf w}_0 \right).
\end{equation}
The eigenvectors (\ref{bounded-H}) and (\ref{genkernel-bounded-H})
are not in $L^2(\mathbb{R},\mathbb{C}^2)$. This fact implies (see
\cite{PKA96}) that adiabatic dynamics of the dark soliton (the
decaying component) induces radiative waves of the continuous
spectrum (the bounded non-decaying component) already at the first
order of the perturbation theory. The coupling between the decaying
and non-decaying components is computed from the dynamical equations
on parameters $s(T)$ and $\theta(T)$.

\subsection{The leading-order frequency of oscillations}

The linear inhomogeneous problem (\ref{first-order-problem}) leads
to a secular growth of $w_1(z,t;T)$ in $t$ unless the
right-hand-side is orthogonal to the kernel of ${\cal H}$. The
orthogonality condition defines a nonlinear equation on parameters
of the dark soliton (\ref{dark-soliton}):
\begin{eqnarray}
\label{constraint}
&&\frac{1}{2} \int_{-\infty}^{\infty} {\rm
sech}^2(\kappa z) {\rm Im}(\partial_T w_0) d (\kappa z) \nonumber \\
&& =\frac{1}{2} \int_{-\infty}^{\infty} {\rm sech}^2(\kappa z) {\rm
Re}(R_1(w_0,\bar{w}_0)) d (\kappa z).
\end{eqnarray}
By computing the integrals directly, we obtain the system of
dynamical equations:
\begin{equation}
\label{particle-motion-nu} \dot{\nu} = (1 - \nu^2) U_{\epsilon}'(s),
\qquad \dot{s} = \nu U_{\epsilon}(s),
\end{equation}
where the last equation is due to the relation
(\ref{scaling-transformation}) between $\nu(T)$ and $v(T) =
\dot{s}$. Closing the system of dynamical equations, we find the
governing equation for the position of the dark soliton:
\begin{equation}
\label{particle-motion} \ddot{s} + V'(s) = 0, \qquad V(s) =
\frac{1}{2} \left( 1 - U_{\epsilon}^2(s) \right).
\end{equation}
The governing equation (\ref{particle-motion}) is equivalent to
the Hamiltonian system of a particle moving in a potential field,
where $V(s)$ stands for the effective potential energy of the
particle.

Using the WKB asymptotic series (\ref{WKB-series}) for
$U^2_{\epsilon}(x)$ and the power approximation
(\ref{first-terms-mu}) for $\mu_{\epsilon}$, we find the power
approximation for the potential function $V(s)$:
\begin{equation}
\label{Taylor-series-W} V(s) = \frac{s^2}{2} + \frac{\epsilon^2
s^2(2 - s^2)}{4 (1 - s^2)^2} + {\rm O}(\epsilon^4),
\end{equation}
where $|s| < 1$. Since the system (\ref{particle-motion}) is valid
at the leading order of the asymptotic series, the dark soliton
oscillates as a harmonic oscillator in the limit $\epsilon \to 0$:
\begin{equation}
\label{oscillator} \ddot{s} + s = 0,
\end{equation}
which is equivalent to the left-hand-side of the main equation
(\ref{pumped-oscillator-main}). The equation (\ref{oscillator}) for
harmonic oscillator is valid in the unit disk (\ref{disc}). Since
the solutions $s(T) = s_0 \cos(T + \delta_0)$ represent circles of
radius $s_0$ on the phase plane $(s,\dot{s})$, the trajectories
remain inside the disk (\ref{disc}) whenever $s_0 < 1$.

We note that the $O(\epsilon^2)$ corrections of the power
approximation (\ref{Taylor-series-W}) are beyond the Tomas-Fermi
approximation (\ref{first-terms-U}). These terms can be dropped in
the main equation (\ref{particle-motion}), since the $O(\epsilon)$
corrections to the main equation (\ref{particle-motion}) occur from
the solution of the second-order inhomogeneous equation
(\ref{second-order-problem}).

\subsection{The first-order radiation corrections}

After the constraint (\ref{constraint}) is added, the linear
inhomogeneous equation (\ref{first-order-problem}) can be solved for
$w_1(z,t;T)$, such that $w_1(z,t;T)$ is bounded in $t$. We adopt a
standard assumption of the asymptotic multi-scale expansion method
that the solution $w_1(z,t;T)$ approaches the stationary solution
$w_{1s}(z;T)$ as $t \to \infty$ due to dispersive decay estimates.
The stationary solution $w_{1s}(z;T)$ can be represented in the
form:
\begin{eqnarray}
\label{first-order-solution}
w_{1s} &=& \frac{q(T)}{U_{\epsilon}^2(s)}
\left( i z w_0 - \partial_{\nu} w_0 \right) \nonumber \\
&+& \frac{3 \nu q(T) - \dot{\theta}(T)}{2 \kappa U_{\epsilon}^2(s)}
\partial_{\kappa} w_0 + \tilde{w}_{1s}(z;T).
\end{eqnarray}
The first two terms in (\ref{first-order-solution}) represent the
linearly growing (\ref{growing-H}) and bounded
(\ref{genkernel-bounded-H}) eigenvectors. The last term
$\tilde{w}_{1s}(z;T)$ is a $t$-independent solution of the
inhomogeneous equation (\ref{first-order-problem}) which corresponds
to the last two terms in the right-hand-side of
(\ref{first-order-problem}) under the constraint
(\ref{particle-motion-nu}). We do not include the decaying
(\ref{kernel-H}) and bounded (\ref{bounded-H}) solutions of the
kernel of ${\cal H}$ in the representation
(\ref{first-order-solution}) as they renormalize parameters
$\theta(T)$ and $s(T)$.

The dependence of $q(T)$ and $\dot{\theta}(T)$ is defined from the
radiation problem, associated to the behavior of the stationary
solution $w_{1s}(z;T)$ as $|z| \to \infty$. Let $w_{1s}^{\pm}(z;T)$
represent a linearly growing solution $w_{1s}(z;T)$ as $z \to \pm
\infty$. Neglecting exponentially small terms in the limits $z \to
\pm \infty$, we obtain from (\ref{first-order-problem}) the linear
inhomogeneous problem for $w^{\pm}_{1s}(z;T)$:
\begin{eqnarray}
\label{first-order-radiation-problem}
&&U_{\epsilon}^2(s) \left[
\frac{1}{2} w_{1s}^{\pm \prime\prime} - i \nu w_{1s}^{\pm \prime} -
w_{1s}^{\pm} - (\kappa \pm i \nu)^2 \bar{w}_{1s}^{\pm} \right]
\nonumber \\
&&= \dot{\theta} ( i \nu \pm \kappa) + \frac{\dot{\nu}}{\kappa} ( \kappa \pm i \nu).
\end{eqnarray}
The most general linearly growing solution of the inhomogeneous
equation (\ref{first-order-radiation-problem}) has the form:
\begin{equation}
\label{radiation-terms} w_{1s}^{\pm} = \frac{1}{U_{\epsilon}^2(s)}
\left[ (a_1 \pm b_1) i z (\pm \kappa + i \nu) + (a_2 \pm b_2)
\right],
\end{equation}
where $a_{1,2}$ and $b_{1,2}$ satisfy two relations:
\begin{equation}
\label{constraint-terms} \nu a_1 - 2 \kappa b_2 = \dot{\theta},
\qquad \nu b_1 - 2 \kappa a_2 = \frac{\dot{\nu}}{\kappa}.
\end{equation}
The first two terms in the solution (\ref{first-order-solution})
have odd real parts and even imaginary parts in $z$, while the
component $\tilde{w}_{1s}(z,T)$ has the opposite symmetry in $z$.
Matching the linear growing terms in (\ref{first-order-solution})
and (\ref{radiation-terms}) under the relation
(\ref{constraint-terms}), we obtain that
\begin{equation}
\label{relation-1-constants} a_1 = q, \qquad b_2 = \frac{\nu q -
\dot{\theta}}{2 \kappa}.
\end{equation}
We note that the constant terms in (\ref{first-order-solution}) and
(\ref{radiation-terms}) are equivalent to the second equation
(\ref{relation-1-constants}) under the renormalization of $\theta =
\theta_0^{\pm} + \epsilon \theta_1^{\pm} + {\rm O}(\epsilon^2)$,
where $\theta_1^{\pm} = \pm q/\kappa$. One more equation is needed
for finding of values of $a_2$ and $b_1$. This equation can be
derived from the balance equation for the renormalized power of the
perturbed NLS equation (\ref{NLS}):
\begin{eqnarray*}
\left( \partial_t - \nu U_{\epsilon}^2(s) \partial_z \right)
n(w,\bar{w}) + U_{\epsilon}^2(s) \partial_z j(w,\bar{w}) =
l(w,\bar{w}),
\end{eqnarray*}
where
\begin{eqnarray}
n &=& |w|^2 - 1, \nonumber \\
j &=&  \frac{1}{2i} \left( \bar{w} w_z - \bar{w}_z w \right), \nonumber \\
l &=& i \left( \bar{w} R(w,\bar{w}) - w \bar{R}(w,\bar{w}) \right).
\end{eqnarray}
Computing explicitly the integral quantities and the jump
conditions:
\begin{eqnarray*}
&&N = \int_{-\infty}^{\infty} n(w,\bar{w}) dz = - 2 \kappa + {\rm O}(\epsilon), \\
&&L = \int_{-\infty}^{\infty} l(w,\bar{w}) dz = 2 \nu \kappa
U'_{\epsilon}(s) \epsilon + {\rm O}(\epsilon^2)
\end{eqnarray*}
and
\begin{eqnarray*}
&&U^2_{\epsilon}(s) \left[ n(w,\bar{w}) \right]_-^+ = 4 \kappa a_2 \epsilon + {\rm O}(\epsilon^2), \nonumber \\
&&U^2_{\epsilon}(s) \left[ j(w,\bar{w}) \right]_-^+ = 2 b_1 \epsilon + {\rm O}(\epsilon^2),
\end{eqnarray*}
we find another relation on $a_2$ and $b_1$:
\begin{equation}
\kappa b_1 - 2 \kappa^2 \nu a_2 = \nu \kappa^2 U_{\epsilon}'(s)  -
\nu \dot{\nu}.
\end{equation}
Using the leading-order equation (\ref{particle-motion-nu}) for
$\dot{\nu}$, we obtain that
\begin{equation}
\label{relation-2-constants} b_1 = 2 \kappa \nu a_2 = -
\frac{\nu}{\kappa} U_{\epsilon}'(s).
\end{equation}
Once the parameters $a_1$, $a_2$, $b_1$, and $b_2$ in the asymptotic
representation (\ref{radiation-terms}) are uniquely found, the
stationary solution $w_s(z;T)$ can be defined outside the dark
soliton up to the order of ${\rm O}(\epsilon^2)$ terms:
\begin{eqnarray*}
\lim_{z \to \pm \infty} w_s(z,T) = \left( 1 + \epsilon W^{\pm}(X,T)
\right) e^{i \Theta^{\pm}(X,T)}\\
= \left[ (\pm \kappa + i \nu) + \epsilon w_{1s}^{\pm}(z;T) + {\rm
O}(\epsilon^2) \right] e^{i \left(\theta_0^{\pm} + \epsilon
\theta_1^{\pm} + {\rm O}(\epsilon^2) \right)},
\end{eqnarray*}
where $X = \epsilon x$, $\epsilon z = U_{\epsilon}(s(T)) (X -
s(T))$,
\begin{eqnarray*}
W^{\pm} & = & \frac{\kappa (b_2 \pm a_2)}{U^2_{\epsilon}(s)} + {\rm
O}(\epsilon), \\
\Theta^{\pm} & = & \Theta_0^{\pm}(T) + \epsilon \Theta_1^{\pm}(T) +
\epsilon z  \frac{(a_1 \pm b_1)}{U^2_{\epsilon}(s)} + {\rm
O}(\epsilon^2),
\end{eqnarray*}
and the explicit forms for $\Theta_0^{\pm}$ and $\Theta_1^{\pm}$ are
not written. The radiation fields $W^{\pm}(X,T)$ and
$\Theta^{\pm}(X,T)$ are defined outside the dark soliton for $X
\gtrless s(T)$, respectively, subject to the boundary conditions:
\begin{eqnarray}
\label{matching-conditions-first} W^{\pm} \biggr|_{X = s(T)} &=&
\frac{\kappa (b_2 \pm a_2)}{U^2_{\epsilon}(s)} + {\rm O}(\epsilon), \nonumber \\
\frac{\partial \Theta}{\partial X} \biggr|_{X = s(T)} &=& \frac{(
a_1 \pm b_1)}{U_{\epsilon}(s)} + {\rm O}(\epsilon).
\end{eqnarray}
These conditions match the inner asymptotic expansion for $z = {\rm
O}(1)$ and the outer asymptotic expansion for $X = {\rm O}(1)$ in
the stationary solution $w_s(z;T)$ (see \cite{PKA96}).

\subsection{Radiation problem for small-amplitude waves}

\begin{figure}[tbp]
\includegraphics[width=8cm]{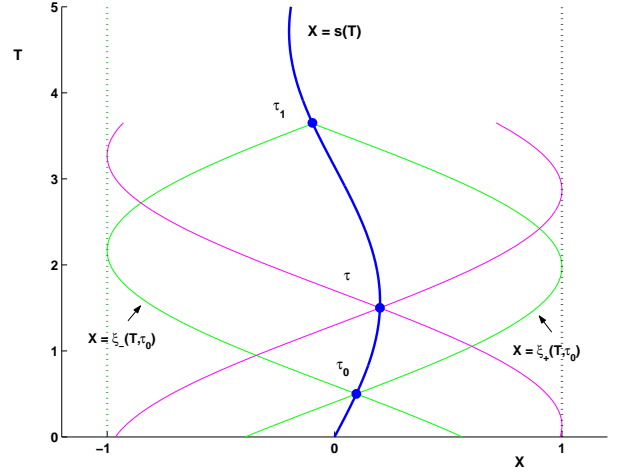}
\caption{The domain of the PDE system
(\ref{Riemann1})--(\ref{Riemann2}) and the family of two
characteristics, superposed with the dark soliton curve $X = s(T)$.}
\label{fig0}
\end{figure}

We now consider the small-amplitude waves within the original
equation (\ref{GP-modified}) for $w(x,t)$. Using the polar form,
$w(x,t) = R(x,t) \exp\left( i \Phi(x,t)\right)$, we obtain the
system:
\begin{eqnarray*}
&&R_t + R_x \Phi_x + \frac{1}{2} R \Phi_{xx} = -
\frac{U_{\epsilon}'(x)}{U_{\epsilon}(x)} R \Phi_x, \\
&&\Phi_t + \frac{1}{2} \Phi_x^2 - \frac{R_{xx}}{2 R} -
U_{\epsilon}^2(x) (1 - R^2)  =
\frac{U_{\epsilon}'(x)}{U_{\epsilon}(x)} \frac{R_x}{R}.
\end{eqnarray*}
The small-amplitude long-wave solutions of the above system
can be constructed in the asymptotic form:
\begin{eqnarray*}
R & = & 1 + \epsilon W(X,T) + {\rm O}(\epsilon^2), \\
\Phi & = & \Theta(X,T) + {\rm O}(\epsilon),
\end{eqnarray*}
where $X = \epsilon x$, $T = \epsilon t$ and we use the fact that $U_{\epsilon}(x) \equiv
U_{\epsilon}(X)$ in the asymptotic region
$|\epsilon x | < 1$ as $\epsilon \to 0$. The leading-order terms
$W(X,T)$ and $\Theta(X,T)$ solve the coupled problem:
\begin{eqnarray}
\label{Riemann1} W_T + \left( U_{\epsilon}(X) V \right)_X + 2
U_{\epsilon}'(X) V & = & 0, \\ \label{Riemann2} \left(
U_{\epsilon}(X) V \right)_T + \left( U_{\epsilon}^2(X) W \right)_X
& = & 0,
\end{eqnarray}
where
\begin{equation}
\label{wave-equation-V} V = \frac{\Theta_X}{2 U_{\epsilon}(X)} .
\end{equation}
Equivalently, the coupled problem (\ref{Riemann1})--(\ref{Riemann2})
with the correspondence (\ref{wave-equation-V}) reduces to the wave equation
with a space-dependent speed:
\begin{equation}
\label{wave-equation} \Theta_{TT} - \left( U_{\epsilon}^2(X)
\Theta_X \right)_X = 0,
\end{equation}
where
\begin{equation}
\label{wave-equation-W} W = - \frac{\Theta_T}{2 U_{\epsilon}^2(X)}.
\end{equation}
The system (\ref{Riemann1})--(\ref{Riemann2}) and the scalar
equation (\ref{wave-equation}) are to be solved separately for $X >
s(T)$ and $X < s(T)$ subject to the boundary conditions
(\ref{matching-conditions-first}). In addition, two radiation
boundary conditions must be added to the system
(\ref{Riemann1})--(\ref{Riemann2}) for a unique solution. Since the
small-amplitude waves move faster than the dark soliton (indeed,
$\dot{s}^2 < U_{\epsilon}^2(s) = 1 - s^2$ in the domain
(\ref{disc})), the radiative waves include the right-travelling wave
for $X > s(T)$ and the left-travelling wave for $X < s(T)$. Figure
\ref{fig0} shows the upper-half $(X,T)$-plane, which is divided by
the curve $X = s(T)$ into two domains $D_{\rm r}$ and $D_{\rm l}$.

The wave equation (\ref{wave-equation}) has two characteristics,
which are defined by the principal part of the PDE system
(\ref{Riemann1})--(\ref{Riemann2}). Let $X = \xi_{\pm}(T)$ be the
equations for the two characteristics curves, starting from a
particular point $(s(\tau_0),\tau_0)$, where $\tau_0 > 0$. According
to the standard text on PDEs \cite{PDEtext}, we find that the
functions $\xi_{\pm}(T;\tau_0)$ solve the initial-value problem for
$T \geq \tau_0$:
\begin{equation}
\label{characteristics} \frac{d \xi_{\pm}}{dT} = \pm
U_{\epsilon}(\xi), \qquad \xi_{\pm}(\tau_0;\tau_0) = s(\tau_0).
\end{equation}
The components $R_{\pm} = W \pm V = R_{\pm}(T;\tau_0)$, defined
along the characteristics $X = \xi_{\pm}(T;\tau_0)$, solve the
system of evolution equations:
\begin{eqnarray}
\label{invariant-1} \frac{d R_+}{d T} & = & - \frac{1}{2}
U'_{\epsilon}(\xi_+(T;\tau_0)) \left( 5 R_+ - R_- \right), \\
\label{invariant-2} \frac{d R_-}{d T} & = & - \frac{1}{2}
U'_{\epsilon}(\xi_-(T;\tau_0)) \left( R_+ - 5 R_- \right).
\end{eqnarray}
Integrating the initial-value problem (\ref{characteristics}) in the
Thomas--Fermi approximation $U_{\epsilon}(X) = \sqrt{1 - X^2}$, we
find the explicit solution:
\begin{equation}
\label{characteristics-invariant} \xi_{\pm}(T;\tau_0) = \sin( \xi_0
\pm (T - \tau_0)),
\end{equation}
where $\xi_0 = \arcsin(s(\tau_0))$. The families of two
characteristics intersect transversely the dark soliton curve $X =
s(T)$ at any point $(s(\tau_0),\tau_0)$ (see Figure \ref{fig0}),
where the components $R_{\pm}$ are generated by means of the
boundary conditions (\ref{matching-conditions-first}) and the
radiation boundary conditions, which need to be added to the system
(\ref{invariant-1})--(\ref{invariant-2}).

Let us consider the domain $D_{\rm r}$ to the right of the curve $X
= s(T)$. By geometry of the initial-value problem (see Figure
\ref{fig0}) or by the symmetry of the coupled problem
(\ref{invariant-1})--(\ref{invariant-2}), the characteristics
$\xi_-(T;\tau_1)$ and the component $R_-(T;\tau_1)$ in $D_{\rm r}$
are the same as the characteristics $\xi_+(T;\tau_0)$ and the
component $R_+(T;\tau_0)$ in $D_{\rm r}$. When integrating the
evolution equation for $R_+(T;\tau_0)$, one needs to substitute the
value for $R_-$, from its value on the transversally intersecting
characteristics in $D_{\rm r}$. Figure 3 shows two intersections of
$\xi_+(T;\tau_0)$ for $\tau_0 < T < \tau_1$ with the characteristics
to the point $(s(\tau),\tau)$. Before the reflection,
$\xi_+(T;\tau_0)$ intersects with $\xi_-(T;\tau)$, such that $R_- =
R_-(T;\tau)$. After the reflection, $\xi_+(T;\tau_0)$ intersects
with $\xi_+(T;\tau)$, such that $R_- = R_+(T;\tau)$. At $T =
\tau_1$, the matching formula gives a required radiation boundary
condition:
\begin{equation}
\label{Sommerfeld-in-fact} R_-(\tau_1;\tau_1) = R_+(\tau_1,\tau_0).
\end{equation}
The initial data for $R_+(\tau_0,\tau_0)$ are uniquely defined from
(\ref{matching-conditions-first}) and (\ref{Sommerfeld-in-fact}).

Similarly, in the domain $D_{\rm l}$, we obtain the radiation
boundary condition:
\begin{equation}
\label{Sommerfeld-in-fact-a} R_+(\tau_1;\tau_1) =
R_-(\tau_1,\tau_0).
\end{equation}
If no incoming waves are imposed initially for $T \leq 0$, then we
obtain that $R_- = 0$ in $D_{\rm r}$ and $R_+ = 0$ in $D_{\rm l}$ at
least for $0 \leq T \leq \tau_*$, where $\tau_*$ is the first
intersection of $\xi_+(T;0)$ with $s(T)$, such that $\xi(\tau_*;0) =
s(\tau_*)$. Therefore, the components $V^{\pm}(X,T)$ and
$W^{\pm}(X,T)$ of the radiative waves to the right and left of the
dark soliton $X = s(T)$ are related at the moving boundary $X =
s(T)$ by the radiation boundary conditions:
\begin{equation}
\label{radiation-boundary-conditions} V^{\pm} \biggr|_{X = s(T)} =
\pm W^{\pm} \biggr|_{X = s(T)}.
\end{equation}
The latter conditions result by virtue of
(\ref{matching-conditions-first}) and (\ref{wave-equation-V}) in the
constraints:
\begin{equation}
a_1 = 2 \kappa a_2, \qquad b_1 = 2 \kappa b_2.
\end{equation}
Using explicit formulas (\ref{relation-1-constants}) and
(\ref{relation-2-constants}), we find unique expressions for $q(T)$
and $\dot{\theta}(T)$:
\begin{equation}
\label{relations-final} q = - \frac{U'_{\epsilon}(s)}{\kappa} ,
\qquad \dot{\theta} = 0.
\end{equation}
Relations
(\ref{radiation-boundary-conditions})--(\ref{relations-final}) are
valid within the first period of oscillations $0 \leq T \leq \tau_*
< 2 \pi$ under the condition that no incoming waves are generated
initially. If absorbing boundary conditions for radiative waves are
specified on the boundary of the TF radius ($X \approx \pm 1$), the
relations
(\ref{radiation-boundary-conditions})--(\ref{relations-final}) are
extended for later times $T > \tau_*$. Otherwise, additional terms
occur in the relations
(\ref{radiation-boundary-conditions})--(\ref{relations-final}) due
to multiple reflections, and these terms are defined by the solution
of the coupled evolution problem
(\ref{invariant-1})--(\ref{invariant-2}) along characteristics
(\ref{characteristics-invariant}) with the radiation boundary
conditions (\ref{Sommerfeld-in-fact})--(\ref{Sommerfeld-in-fact-a}).

\subsection{The first-order corrections to the main equation}

Radiative corrections to the main equation for $s(T)$ are derived
from the second-order inhomogeneous equation
(\ref{second-order-problem}) by means of the same orthogonality
condition (\ref{constraint}). Equivalently, radiative corrections
can be computed from the balance equation for the renormalized
momentum \cite{PKA96}:
\begin{eqnarray}
&&\left( \partial_t - \nu U_{\epsilon}^2(s)
\partial_z \right) p(w,\bar{w}) + U^2_{\epsilon}(s) \partial_z
r(w,\bar{w}) \nonumber \\ \label{momentum-balance} &&
\phantom{texttext}  = \partial_z m(w,\bar{w}) + k(w,\bar{w}),
\end{eqnarray}
where
\begin{eqnarray*}
&& p = \frac{i}{2} \left( \bar{w} w_z - w \bar{w}_z \right)
\left( 1 - \frac{1}{|w|^2} \right), \\
&& r = \frac{1}{4} \left( \bar{w} w_{zz} - 2 \bar{w}_z w_z + w
\bar{w}_{zz} \right) \left( 1 - \frac{1}{|w|^2} \right) \\
&& \phantom{texttexttext} - \frac{|w_z|^2}{2 |w|^2} - \frac{1}{2}
\left( 1 - |w|^2 \right)^2
\end{eqnarray*}
and
\begin{eqnarray*}
&& m = -\frac{1}{2} \left( \bar{w} R(w,\bar{w}) +
w \bar{R}(w,\bar{w}) \right) \left( 1 - \frac{1}{|w|^2} \right), \\
&& k = \bar{w}_z R(w,\bar{w}) + w_z \bar{R}(w,\bar{w}).
\end{eqnarray*}
Expanding the integral quantities and the jump conditions into the
asymptotic approximations:
\begin{eqnarray*}
P & = & \int_{-\infty}^{\infty} p(w,\bar{w}) dz = P_0 + \epsilon
P_1 + {\rm O}(\epsilon^2), \\
K & = & \int_{-\infty}^{\infty} k(w,\bar{w}) dz = \epsilon K_1 +
\epsilon^2 K_2 + {\rm O}(\epsilon^3)
\end{eqnarray*}
and
\begin{eqnarray*}
U^4_{\epsilon}(s) [p]^+_- & = & - 4 \epsilon^2 \kappa (a_1
a_2 + b_1 b_2) + {\rm O}(\epsilon^3), \\
U^4_{\epsilon}(s) [r]^+_- & = & - 2 \epsilon^2 (a_1 b_1 + 4
\kappa^2 a_2 b_2) + {\rm O}(\epsilon^3) \\
U^4_{\epsilon}(s) [m]^+_- & = & {\rm O}(\epsilon^3),
\end{eqnarray*}
we find the explicit expressions
\begin{eqnarray*}
&& P_0 = 2 \nu \kappa - 2 \arctan \left(
\frac{\kappa}{\nu}\right), \\
&& U^2_{\epsilon}(s) P_1 = 2 \kappa q  - q
\partial_{\nu} P_0 + \frac{3 \nu q - \dot{\theta}}{2 \kappa}
\partial_{\kappa} P_0 = 2 U'_{\epsilon}(s), \\
&& K_1 = 4 \kappa (1 - \nu^2) U'_{\epsilon}(s),\\
&& U^2_{\epsilon}(s) K_2 = - 8 \nu \kappa q U'_{\epsilon}(s) - q
\partial_{\nu} K_1 + \frac{3\nu q - \dot{\theta}}{2 \kappa}
\partial_{\kappa} K_1 \\
&& \phantom{texttextte} = -6 \nu (U'_{\epsilon}(s))^2
\end{eqnarray*}
and
$$
U^4_{\epsilon}(s) [- \nu p + r]^+_- = -2 \epsilon^2 \nu \left(
U_{\epsilon}'(s)\right)^2 + {\rm O}(\epsilon^3).
$$
We note that the expressions for $P_0$ and $K_1$ are found by using
$w_0(z)$ in Eq. (\ref{dark-soliton}) as a function of two
independent parameters $\kappa$ and $\nu$. This technical trick
allows us to compute the correction terms $P_1$ and $K_2$ by using
partial derivatives of $P_0$ and $K_1$ in $\kappa$ and $\nu$. The
final expressions use the relation $\kappa = \sqrt{1 - \nu^2}$ as
well as the previously found relations (\ref{relation-1-constants}),
(\ref{relation-2-constants}), and (\ref{relations-final}).

Substituting all expansions into the balance equation
(\ref{momentum-balance}) and integrating on $z \in \mathbb{R}$, we
obtain an extended main equation in the form:
$$
\dot{\nu} - \kappa^2 U_{\epsilon}'(s) + \frac{\epsilon \nu
U_{\epsilon}''(s)}{2 \kappa U_{\epsilon}(s)} = {\rm O}(\epsilon^2).
$$
Using the relation $\dot{s} = \nu U_{\epsilon}(s)$, the main
equation can be rewritten for $s(T)$:
\begin{equation}
\label{extended-equation} \ddot{s} - U_{\epsilon}(s)
U_{\epsilon}'(s) + \frac{\epsilon \dot{s} U_{\epsilon}''(s)}{2
\kappa U_{\epsilon}(s)} = {\rm O}(\epsilon^2).
\end{equation}
Furthermore, using the Tomas-Fermi approximation $U_{\epsilon}(s) =
\sqrt{1 - s^2}$, the extended dynamical equation is rewritten in the
final form:
\begin{equation}
\label{pumped-oscillator} \ddot{s} + s = \frac{\epsilon \dot{s}}{2
\kappa (1-s^2)^2} + {\rm O}(\epsilon^2),
\end{equation}
where
$$
\kappa = \sqrt{\frac{1 - s^2 - \dot{s}^2}{1 - s^2}}
$$
and $s^2 + \dot{s}^2 < 1$. We note that the perturbation term $R_2$
in the right-hand-side of the second-order problem
(\ref{second-order-problem}) does not contribute to the main
equation (\ref{pumped-oscillator}), since all associated integrals
in the correction term $K_2$ are zero due to symmetry in $z \in
\mathbb{R}$.

Second-order corrections ${\rm O}(\epsilon^2)$ to the extended
equation (\ref{pumped-oscillator}) can be incorporated to the
asymptotic theory if the second-order problem
(\ref{second-order-problem}) is solved explicitly and the
third-order problem associated to the perturbed NLS equation
(\ref{NLS}) is analyzed. It is beyond the scope of our manuscript to
derive the error bounds between the solution of the perturbed NLS
equation (\ref{NLS}) and the truncated stationary solution of the
first-order and second-order problems (\ref{first-order-problem})
and (\ref{second-order-problem}).

\section{Failure of the formal perturbation theory}

We shall consider the perturbed GP equation in the form
(\ref{GP-modified}), where the ground state $U_{\epsilon}(x)$ is
truncated by the Thomas-Fermi approximation (\ref{first-terms-U}),
such that $U_{\epsilon} = \sqrt{1 - \epsilon^2 x^2}$. The perturbed
GP equation (\ref{GP-modified}) takes the explicit form:
\begin{equation}
\label{GP-formal}
 i w_t + \frac{1}{2} w_{xx} + (1 - |w|^2) w = R(w,\bar{w}),
\end{equation}
where
$$
R(w,\bar{w}) = \epsilon^2 x^2 (1 - |w|^2) w + \frac{\epsilon^2 x}{1
- \epsilon^2 x^2} w_x.
$$
The GP equation (\ref{GP-formal}) has the exact solution for
$R(w,\bar{w})= 0$:
\begin{equation}
\label{dark-formal} w(x,t) = w_0(\eta) = k \tanh(k \eta) + i v,
\end{equation}
where
\begin{equation}
\label{scaling-formal} \eta = x - \frac{s(T)}{\epsilon}, \qquad v(T)
= \dot{s}, \qquad T = \epsilon t.
\end{equation}
If $w_0(\eta)$ is a steadily traveling dark soliton, parameters are
constant, where $k = \sqrt{1- v^2}$ is the amplitude and $v$ is the
speed. We assume that the coordinate $s(T)$ of the dark soliton
changes adiabatically under the small perturbation $R(w,\bar{w})
\neq 0$ and show that a formal perturbation theory fails to capture
the correct dependence of the frequency of oscillations. Using the
same renormalized momentum equation (\ref{momentum-balance}), we
obtain from (\ref{GP-formal}) the leading-order balance equation for
the renormalized momentum of the dark soliton \cite{KY94,CCH98,L04}:
\begin{equation}
\label{momentum-formal} \epsilon \frac{d P_s}{d T} = -
\int_{-\infty}^{\infty} w_0'(x) \left( R + \bar{R}
\right)(w_0,\bar{w}_0) dx,
\end{equation}
where
\begin{eqnarray}
P_s &=& \frac{i}{2} \int_{-\infty}^{\infty} \left( \bar{w}_0 w_0' - w
\bar{w}_0' \right) \left( 1 - \frac{1}{|w_0|^2} \right) dx \nonumber \\
&=& 2 v k - 2 {\rm arctan} \left( \frac{k}{v} \right),
\end{eqnarray}
such that
$$
\epsilon \frac{d P_s}{d T} = -\frac{4 \epsilon k^2
\dot{k}}{\sqrt{1-k^2}}.
$$
The same equation occurs in the Lagrangian averaging technique
applied to the perturbed GP equation (\ref{GP-formal}) (see
\cite{BP05} and references therein). The integrals in the
right-hand-side of (\ref{momentum-formal}) can be evaluated at the
leading-order approximation as $\epsilon \to 0$, when the scaling
(\ref{scaling-formal}) is used. As a result, we have
\begin{eqnarray*}
&& \epsilon^2 \int_{-\infty}^{\infty} x^2 (1 - |w_0|^2) (|w_0|^2)'
dx = \frac{4}{3} \epsilon k^3 s + {\rm O}(\epsilon^3), \nonumber \\
&& 2 \epsilon^2 \int_{-\infty}^{\infty} \frac{x |w_0'|^2}{1-
\epsilon^2 x^2} dx  =  \frac{8}{3} \frac{\epsilon k^3 s}{1-s^2}
+{\rm O}(\epsilon^3).
\end{eqnarray*}
Using the leading-order approximation $\dot{s} = \sqrt{1 - k^2}$, we
close the main equation for $s(T)$ as follows:
\begin{equation}
\label{main-formal}
\ddot{s} + \frac{(3 - s^2)(1-\dot{s}^2)}{3 (1 - s^2)} s = {\rm O}(\epsilon^2).
\end{equation}
The main equation (\ref{main-formal}) represents the adiabatic
approximation for dynamics of dark solitons, in neglecting of the
radiative effects (see \cite{L04,BP05}). In the limit of black
soliton, when $s$ is small, the equation for an anharmonic
oscillator (\ref{main-formal}) approaches the equation for a
harmonic oscillator (\ref{oscillator}). However, for a dark soliton
of arbitrary amplitude, the anharmonic oscillator equation
(\ref{main-formal}) is different from the harmonic oscillator
equation (\ref{oscillator}), although it represents the same
asymptotic limit of adiabatic oscillations of the dark soliton.

The paradox above has a simple resolution. The first term in the
right-hand-side of the perturbed GP equation (\ref{GP-formal})  is
not a small perturbation for the dark soliton (\ref{dark-formal})
under the scaling (\ref{scaling-formal}) in the limit $\epsilon \to
0$. Indeed, $\epsilon^2 x^2 = s^2 - 2 \epsilon s \eta + \epsilon^2
\eta^2$, such that the correct perturbed GP equation in the variables
(\ref{scaling-formal}) takes the form:
\begin{equation}
\label{GP-formal-accurate} i w_t - i v w_{\eta} + \frac{1}{2}
w_{\eta \eta} + (1-s^2) (1 - |w|^2) w = \tilde{R}(w,\bar{w}),
\end{equation}
where
$$
\tilde{R} = -2 \epsilon s \eta (1 - |w|^2) w + \frac{\epsilon s}{1 -
s^2} w_{\eta} + {\rm O}(\epsilon^2).
$$
However, the dark soliton $w = w_0(\eta)$ is no longer a solution of
the left-hand-side of the GP equation (\ref{GP-formal-accurate}) and
the renormalization of the variable $\eta$ is required before formal
application of the perturbation theory. The renormalization of the
GP equation is developed in the main part of this paper. We conclude
that the formal application of the perturbation theory to the
perturbed GP equation (\ref{GP-formal}) fails to recover an accurate
dependence of frequency of dark soliton oscillations from the
amplitude of the dark soliton.

\section{Conclusions}

In this paper, we have analyzed the oscillations of dark solitons in
trapped atomic Bose-Einstein condensates. We have considered a
repulsive quasi-one-dimensional condensate, described by a
Gross-Pitaevskii equation with a parabolic external trapping
potential. An asymptotic multi-scale expansion method has been
developed in the limit of the weakly trapped condensate. To the
leading-order of approximation (i.e., neglecting the
inhomogeneity-induced sound emission by the soliton), the soliton
motion is harmonic, with a constant frequency depending on the trap
strength. This result bridges earlier predictions obtained by
different approaches in two limiting cases, namely the adiabatic
perturbation theory for nearly black solitons \cite{FT02} and the
Korteweg-deVries approximation for shallow solitons \cite{HH02}, and
is also in agreement with the recent prediction based on the
semi-classical Landau dynamics of the dark soliton \cite{KP04}. On
the other hand, to first-order approximation, the radiation (sound
waves) emitted by the soliton due to the inhomogeneous background is
taken into account. It is shown that radiation plays an important
role in the soliton motion, as it responsible to an anti-damping
effect, resulting in the increase of the amplitude of oscillations
and decrease of the soliton amplitude. Energy loss of the soliton
due to the emission of radiation is approximately found to follow an
acceleration-square law, in agreement with previous numerical
observations \cite{prok1}. We have also compared the results of the
presented multi-scale expansion technique with the formal
perturbation theory for dark solitons. We have found that a formal
application of the perturbation theory fails to incorporate the
correct dependence of the frequency of the dark soliton oscillation
on the soliton amplitude (velocity). Numerical results have been
found to be in good agreement to the analytical predictions, within
the applicability intervals of the analytical assumptions. Finally,
it should be noticed that the presented technique can also be used
for the study of dark soliton dynamics in other relevant
(inhomogeneous) systems, such as optical lattices and superlattices.

\vspace{5mm}

{\bf Acknowledgements} This work was partially supported by
NSF-DMS-0204585, NSF-CAREER, and the Eppley Foundation for Research
(PGK) and by NSERC grant (DEP).

\end{document}